\documentclass[aps,pre,superscriptaddress,notitlepage,twocolumn]{revtex4-1}
\usepackage{hyperref}
\usepackage{amsmath}
\usepackage{graphicx}
\usepackage{color}

\newcommand{\be}{\begin{equation}}
\newcommand{\ee}{\end{equation}}

\newcommand{\bs}{\boldsymbol}

\newcommand{\ket}[1]{\left|#1\right>}      
\newcommand{\bra}[1]{\left<#1\right|}

\newcommand{\pcsadd}{Center for Theoretical Physics of Complex Systems, Institute for Basic Science, Daejeon 34126, Korea}
\newcommand{\ustadd}{Basic Science Program, Korea University of Science and Technology, Daejeon 34113, Korea}

\begin{document}

\title{Probing bulk topological invariants using leaky photonic lattices}
\author{Daniel Leykam}
\affiliation{\pcsadd}
\affiliation{\ustadd}

\author{Daria A. Smirnova}
\affiliation{Nonlinear Physics Centre, Australian National University, Canberra ACT 2601, Australia}
\affiliation{Institute of Applied Physics, Russian Academy of Science, Nizhny Novgorod 603950, Russia}

\date{\today}

\begin{abstract}
Topological invariants characterising filled Bloch bands attract enormous interest, underpinning electronic topological insulators and analogous artificial lattices for Bose-Einstein condensates, photons, and acoustic waves. In the latter bosonic systems there is no Fermi exclusion principle to enforce uniform band filling, which makes measurement of their bulk topological invariants challenging. Here we show how to achieve controllable filling of bosonic bands using leaky photonic lattices. Leaky photonic lattices host transitions between bound and radiative modes at a critical energy, which plays a role analogous to the electronic Fermi level. Tuning this effective Fermi level into a band gap results in disorder-robust dynamical quantization of bulk topological invariants such as the Chern number. Our findings establish leaky lattices as a novel and highly flexible platform for exploring topological and non-Hermitian wave physics.
\end{abstract}

\maketitle

\section{Introduction}

Characterisation of topological phases has attracted broad interest throughout physics since the discovery of topological insulators described by quantised topological invariants of filled electronic bands~\cite{topo_review}. Demonstrations of topological phases have spread from electronic condensed matter to bosonic systems such as photonics~\cite{topological_photonics_review}, Bose-Einstein condensates~\cite{BEC_review}, and acoustics~\cite{topo_acoustic_review}, where there is no Fermi exclusion principle and hence no concept of band filling. The lack of band filling complicates measurement of bulk topological invariants, demanding indirect approaches based on observing their protected edge states~\cite{mittal2016,chong}, or time-consuming methods based on Bloch band tomography or adiabatic transport~\cite{bardyn2014,aidelsburger2015,wimmer2017,tarnowski2019}. Here we propose a platform that overcomes these challenges, enabling the controlled ``filling'' of bosonic Bloch bands, direct bulk measurements of their topological invariants, and novel analogies between electronic condensed matter and classical wave systems.

We consider wave propagation in shallow lattices supporting a mixture of bound and leaky modes~\cite{quasinormal_review,leaky_review,leaky_review_2,leaky_variational,powell,pick}. Leaky modes, also known as quasi-normal modes, emerge when the effective coupling strength between different lattice sites exceeds the site potential depth, resulting in modal energy-dependent radiative losses. While quasi-normal modes have been extensively employed in the modelling of open scattering systems~\cite{quasinormal_review_2}, their dynamical properties have received comparatively little attention, particularly in the context of topological phases. 

\begin{figure}

\includegraphics[width=\columnwidth]{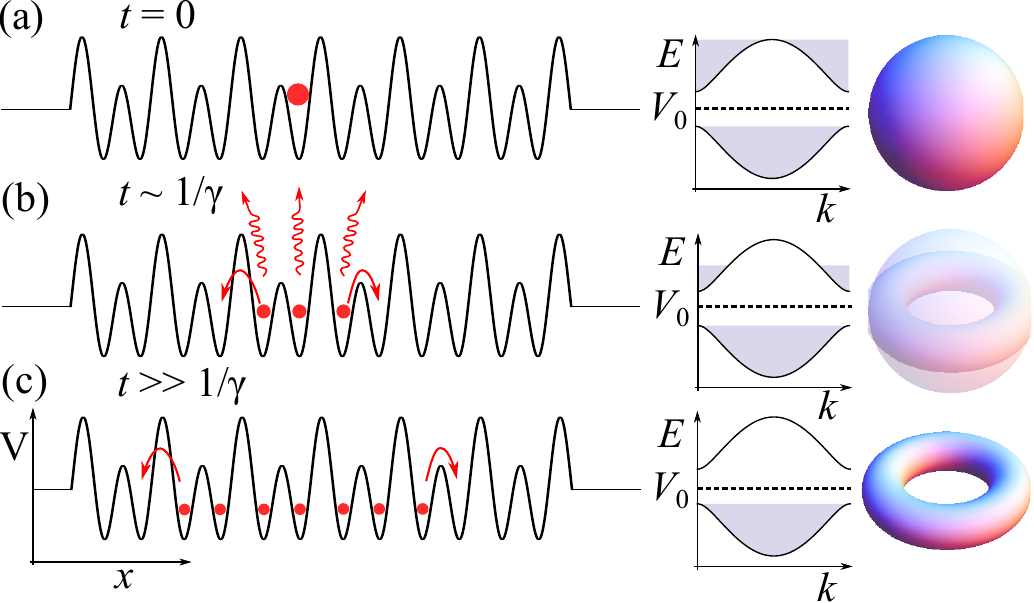}

\caption{Schematic of method to measure topological invariants using leaky lattices. (a) A spatially-localised excitation (red circle) excites all Bloch bands of the lattice and is hence insensitive to the individual bands' topology. (b) Dynamics: The excitation diffracts through the lattice, while quasi-normal modes of Bloch bands with energies $E > V_0$ have a finite lifetime and radiate their energy. (c) At times exceeding the leaky mode lifetime $1/\gamma$ only bound bands remain populated and their topology is imprinted on time-independent observables of the (still-diffracting) field.}

\label{fig:schematic}

\end{figure}

The cutoff energy between bound and leaky quasi-normal modes is analogous to the electronic Fermi level; all modes above this energy dynamically decay, resembling a form of evaporative cooling. By controlling the cutoff energy one can achieve controlled filling of a desired number of bands, as illustrated schematically in Fig.~\ref{fig:schematic}. In effect, an arbitrary initial field profile is projected onto the bound bands, which we will show enables direct measurement of bulk topological invariants such as the Zak phase and Chern number~\cite{discretised,bianco2011,ringel2011,review}. Additionally, this class of shallow lattices provides a novel platform for exploring non-Hermitian tight binding models and topological phases with energy-dependent losses~\cite{lattice_loss,longhi1,longhi2, leykam2017,mukherjee,farfield,non-Hermitian_topology,PT_review,hatsugai}. We will explain and validate our proposal using coupled mode theory for the Su-Schrieffer-Heeger and Haldane models, and full wave simulations of leaky optical waveguide arrays.

This article is structured as follows: In Sec.~\ref{sec:shallow} we briefly review the properties of quasi-normal modes and coupled mode theories for shallow lattices, then show in Sec.~\ref{sec:projector} how their propagation dynamics can be used to obtain band projection operators. Sec.~\ref{sec:SSH} applies this general formalism to the Su-Schrieffer-Heeger model, providing a simple and disorder-robust way to measure its quantised Zak phase. Notably, our approach is superior to existing dynamical approaches such as measuring the mean wavepacket displacement~\cite{rudner, PT_lattice, quantum_SSH}, which is sensitive to disorder. Sec.~\ref{sec:haldane} generalizes to measurement of the Chern number in the two-dimensional Haldane model, showing that the topological invariant can be faithfully measured even in relatively small lattices. In Sec.~\ref{sec:slab} we validate our approach by carrying out numerical simulations of a slab waveguide array implementing a leaky Su-Schrieffer-Heeger model. Sec.~\ref{sec:conclusion} concludes the article with a summary and discussion of future directions and possible experiments.

\section{Wave Propagation In Leaky Lattices}
\label{sec:shallow}

Many open linear wave systems can be described by the Schr\"odinger equation
\be
i \partial_t \psi (\bs{r},t) = [-\frac{1}{2m}\nabla^2 + V(\bs{r})] \psi(\bs{r},t), \label{eq:3DSE}
\ee
where $t$ is the evolution time, $m$ is the wave effective mass, and $V(\bs{r})$ is a localised potential profile. For example, Eq.~\eqref{eq:3DSE} is equivalent to the paraxial equation under the replacements $t \rightarrow z$, $\bs{r} = (x,y)$, $m \rightarrow k_0$, and $V(\bs{r}) \rightarrow - k_0 \delta n(\bs{r})/n_0$, where $z$ is the propagation distance, $k_0 = 2 \pi n_0/\lambda$ is the wavenumber, $n_0$ the ambient refractive index, $\lambda$ is the free space wavelength, and $\delta n (x)$ is the deviation of the refractive index from $n_0$. 

Propagation-invariant modes of Eq.~\eqref{eq:3DSE} with time dependence $\psi(\bs{r},t) = \phi(\bs{r}) e^{-i (E+i\gamma)t}$ satisfy
\be 
\hat{H} \phi = \left[ -\frac{1}{2m} \nabla^2 + V(\bs{r}) \right] \phi_j = (E + i \gamma) \phi, \label{eq:helmholtz}
\ee
where $E$ is the mode energy (propagation constant in the case of waveguide arrays), and $\gamma$ is its growth rate, which can be nonzero in the presence of gain or loss. Often the full propagation dynamics are well-approximated by expanding the continuous wave field amplitude $\psi(\bs{r},t)$ as a superposition of a discrete set of the modes,
\be 
 \psi(\bs{r}, t) = \sum_j c_j  \phi_j(\bs{r}) e^{-i E_j t + \gamma_j t},
\ee
where $c_j$ is the amplitude of the $j$th mode.

When $V(\bs{r})$ describes a lattice of weakly coupled potential wells it is useful to obtain further analytical insight by approximating the system with an effective tight binding Hamiltonian, yielding the discretised eigenvalue problem~\cite{topological_photonics_review,BEC_review}
\be 
\hat{H} = \sum_{n,m} \hat{a}_m^{\dagger} H_{mn} \hat{a}_n, \quad \hat{H} \mid \phi_j \rangle = (E_j + i \gamma_j) \mid \phi_j \rangle, \label{eq:regular}
\ee
where $\hat{a}_n^{\dagger}$ is the creation operator for the field at the $n$th lattice site and for later convenience we have introduced the bra-ket notation to represent wave fields on a discrete lattice. Diagonal elements of $\hat{H}$ describe the energies of the individual lattice sites, while off-diagonal elements describe the coupling between them. In the absence of gain or loss, $\hat{H}$ is Hermitian and its eigenvalues are purely real. One can make $\hat{H}$ non-Hermitian by introducing gain or loss, e.g. via absorption or modal symmetry-dependent radiative losses~\cite{farfield}. Such non-Hermitian Hamiltonians attract enormous interest nowadays~\cite{PT_review}. 

Regardless of whether $\hat{H}$ is Hermitian or non-Hermitian, applying a uniform shift to the energies of the tight binding Hamiltonian (i.e. making the replacement $\hat{H} \rightarrow \hat{H} - V_0 \hat{1}$) does not affect the dynamics; it only introduces an irrelevant phase shift. This is because the usual tight binding approximation assumes a deep lattice, such that each site hosts a bound mode which does not radiate energy to its environment. On the other hand, in shallow (leaky) lattices the number of bound modes may be less than the number of lattice sites, and the detuning of the site energies with respect to the energy of their environment $V_0$ plays a critical role, determining the transition between bound and radiative modes.

When there is a sharp boundary between the lattice and its environment and no backscattering of waves from the environment, one can introduce quasi-normal modes to describe wave propagation dynamics in shallow lattices~\cite{quasinormal_review,leaky_review,leaky_review_2,leaky_variational,powell,pick}. Quasi-normal modes are calculated by eliminating the continuum of radiation modes to obtain an eigenvalue problem on a finite interval with outgoing wave boundary conditions. Because the radiation rate depends on the mode energy, this eigenvalue problem is in general nonlinear. For simplicity, in the following we will assume that radiation occurs via independent, quasi-1D channels described by quadratic dispersion relations of the form $E(k) = V_0 + k^2/(2m)$. Then the resulting quadratic eigenvalue problem can be linearised as~\cite{QEP}
\be 
\left(\begin{array}{cc} \hat{\Gamma} & \hat{H} - V_0 \hat{1} \\ -\hat{1} & 0 \end{array} \right) \mid \Phi_j \rangle = -\xi \mid \Phi_j \rangle, \label{eq:leakyH}
\ee
where $\xi = \sqrt{2m(V_0 - E_j - i \gamma_j)}$ is the field decay rate in the environment, $\mid \Phi_j \rangle = (\xi \mid \phi_j \rangle, \mid \phi_j \rangle )^T$, and $\hat{\Gamma}$ describes the coupling of the individual sites to the environment. When $\hat{H}$ is Hermitian the modes of Eq.~\eqref{eq:leakyH} have purely real eigenvalues for $E_j < 0$, and only acquire complex eigenvalues for $E_j > V_0$~\cite{quasinormal_review} (see Appendix~\ref{sec:continuum} for further discussion).

The transition between real and complex eigenvalues corresponds to a non-Hermitian degeneracy at which incoming and radiating quasi-normal modes coalesce. Such energy-dependent losses provide a novel kind of non-Hermitian propagation dynamics, in contrast to Eq.~\eqref{eq:regular}, where non-Hermitian perturbations generally give all eigenvalues nonzero imaginary parts in the absence of any special symmetries (e.g. parity-time symmetry~\cite{PT_review}). In particular, waves propagating in a leaky lattice will radiate energy from all Bloch modes with energies $E > V_0$, while conserving the population of modes with $E < V_0$. 

\begin{figure}

\includegraphics[width=\columnwidth]{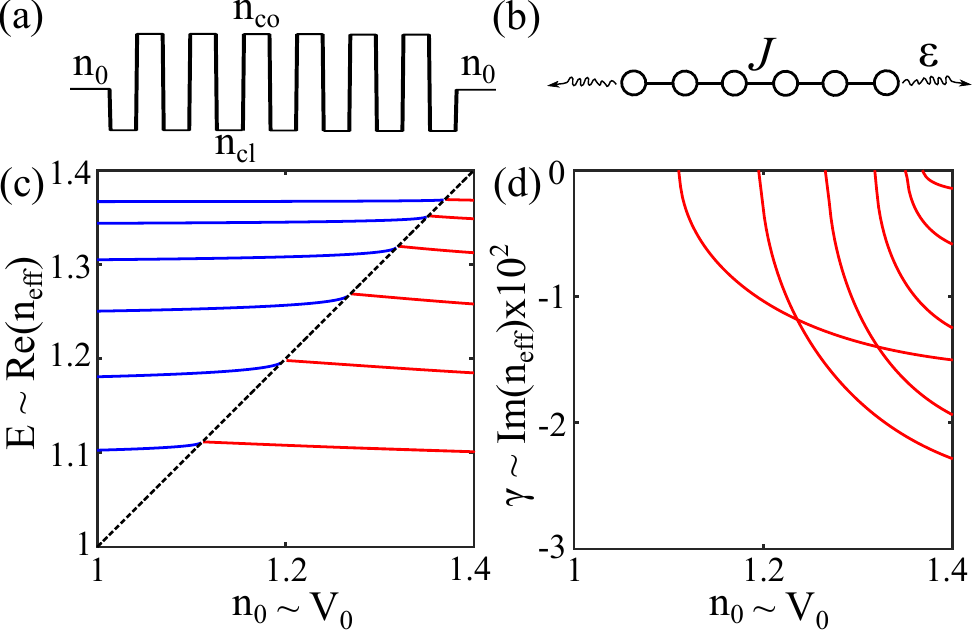}
    \caption{Quasi-normal mode spectrum of a one-dimensional slab waveguide array. (a) Refractive index profile. (b) Corresponding tight binding model, with inter-site coupling $J$ and environmental coupling $\varepsilon$. (c) Modal energies; purely real eigenvalues plotted in blue, while leaky modes shown in red emerge when $E$ crosses $V_0$ (black dashed line). (d) Growth rates of the leaky modes.}
    \label{fig:example}
\end{figure}

As an example of this sharp transition between bound and leaky modes, Fig.~\ref{fig:example} illustrates the spectrum of a one-dimensional slab waveguide array. We obtain its transverse electric (TE) modes by solving the Helmholtz Eq.~\eqref{eq:helmholtz} with $V(x) = -n^2(x)$, $m = k_0^2/2$, and $E-i\gamma = -n_{\mathrm{eff}}^2$, where $n(x)$ is the refractive index profile shown in Fig.~\ref{fig:example}(a) and $n_{\mathrm{eff}}$ is the modal effective index. As the environmental potential depth $V_0$ is decreased ($n_0$ increased), the array modes become leaky one by one, each exhibiting a sharp transition from real to complex eigenvalues. The loss of the leaky modes is determined by both their energy, and how strongly their spatial profile overlaps with the boundaries of the array. We find good qualitative agreement between the numerical solution of the Helmholtz equation and an effective tight binding model of the form $H_{mn} = J(\delta_{n,m-1} + \delta_{n,m+1})$, with $\Gamma_{mn} = \varepsilon \delta_{mn} (\delta_{m,1} + \delta_{m,N})$, illustrated schematically in Fig.~\ref{fig:example}(b).

\section{Band projection using leaky lattices}
\label{sec:projector}

It is particularly interesting to apply the above quasi-normal mode formalism to multi-band lattices, because tuning $V_0$ into a band gap allows one to filter out higher band components of any initial excitation without requiring prior knowledge of their Bloch functions, which we will now show enables direct measurement of band projection operators $\hat{P} = \sum_{E_j < V_0} \mid \phi_j \rangle \langle \phi_j \mid$ and their topological invariants. 

Consider the evolution of an arbitrary excitation of a translation invariant leaky lattice, for which the modes are Bloch functions $\mid u_n(\bs{k}) \rangle$, where $\bs{k}$ is the Bloch momentum, $n$ is the band index, and we henceforth use bra-ket notation to encode internal (sublattice or spin-like) degrees of freedom within each unit cell. The Fourier transform of the field $\mid \psi(\bs{k},t) \rangle$ can be expressed as a superposition of modes from different bands,
\be 
\mid \psi (\bs{k}, t) \rangle = \sum_{n} c_n(\bs{k}) e^{-i [E_n(\bs{k}) + i \gamma_n(\bs{k})] t} \mid u_n (\bs{k}) \rangle, \label{eq:f1}
\ee
where $c_n(\bs{k}) = \langle u_n (\bs{k}) \mid \psi(\bs{k}, 0) \rangle$ are the Bloch function weights, determined by the initial field profile.

The band projection operator $\hat{P}(\bs{k})$ can be obtained by introducing the field projection operator,
\be 
\hat{F}(\bs{k},t) = \mid \psi (\bs{k},t) \rangle \langle \psi (\bs{k},t) \mid.
\ee
Expanding $\hat{F}(\bs{k},t)$ using Eq.~\eqref{eq:f1} (and dropping $\bs{k}$ arguments for brevity),
\be 
\hat{F}(t)  = \sum_{m,n}  c_n c_m^* e^{i (E_m-E_n)t + (\gamma_m + \gamma_n)t } \mid u_n \rangle \langle u_m \mid .
\ee
Leaky modes with energies $E_n > V_0$ above the cutoff will have $\gamma_n <0$, such that their weights become exponentially small at large $t$. Thus, the summation over all bands can be replaced with summation over bands below the cutoff, which have purely real eigenvalues,
\be 
\hat{F}(t)  = \sum_{E_{m,n}<V_0}  c_n c_m^* e^{i (E_m-E_n)t} \mid u_n \rangle \langle u_m \mid .
\ee
The only remaining time-dependent terms are interband terms with $m \ne n$. The interband terms can be eliminated by either measuring $\hat{F}(t)$ as either a time average, or in real space (see Appendix~\ref{sec:phase}). In the latter case interband terms decay at least as fast as $\propto \sqrt{1/t}$. Thus, after sufficiently long $t$ the only terms contributing significantly to the field projector are diagonal in the band index, 
\be 
\hat{F}(t \rightarrow \infty)  = \sum_{E_n < V_0}  |c_n(\bs{k})|^2 \mid u_n \rangle \langle u_n \mid.
\ee
This is almost a band projection operator, spoiled only by the $\bs{k}$-dependence of the band populations, $|c_n(\bs{k})|^2 \ne 1$. To obtain the band projection operator one can make a series of measurements using a set of initial states $\mid \psi^{(j)}(\bs{k},0) \rangle$ localised to a single unit cell (therefore having a flat spectrum in $\bs{k}$ space) and spanning the local Hilbert space of the internal (sublattice or spin) degree of freedom. For example, in a waveguide lattice one should excite in turn each sublattice of a single unit cell. This approach is similar in spirit to Brillouin zone spectroscopy~\cite{bartal2005}. Since the Bloch functions form a complete basis for the local Hilbert space at each $\bs{k}$, the weights of this spanning set satisfy $\sum_{j} |c_n^{(j)}(\bs{k})|^2 = 1$. Consequently, the averaged field projection operator
\be 
\sum_j \hat{F}^{(j)}(\bs{k}) = \sum_{E_n < V_0} \mid u_n (\bs{k}) \rangle \langle u_n (\bs{k}) \mid \; \equiv \hat{P}(\bs{k}), \label{eq:projector_measurement}
\ee
can be used to measure the band projection operator $\hat{P}(\bs{k})$. This is our central result. In the following Sections we will show how topological invariants can be robustly extracted by computing $\hat{P}(\bs{k})$ using the field projection operators $\hat{F}^{(j)}(\bs{k})$, which can be readily obtained by measuring the field amplitude at a fixed time after the leaky modes have decayed.

\section{Zak phase in the Su-Schrieffer-Heeger model}
\label{sec:SSH}

As a first example of measuring topological invariants using leaky photonic lattices, we consider the Su-Schrieffer-Heeger (SSH) model described by the Hamiltonian~\cite{malkova2009},
\be 
\hat{H} = \sum_{n=1}^L \left[ J_1 \hat{a}_n^{\dagger} \hat{b}_{n} +  J_2 \hat{a}_{n+1}^{\dagger} \hat{b}_{n} \right]+ \text{h.c.}
\ee
where $\hat{a}_n^{\dagger}$ ($\hat{b}_n^{\dagger}$) creates a particle on the $a$ ($b$) sublattice in unit cell $n$, $J_1$ and $J_2$ are intra- and inter-cell hopping strengths, respectively, and the lattice consists of $L$ unit cells. The Bloch Hamiltonian describing the system under periodic boundary conditions is
\be 
\hat{H}(k) = \left( \begin{array}{cc} 0 & J_1 + J_2 e^{ik} \\ J_1  + J_2 e^{-ik} & 0 \end{array} \right).
\ee
Because of the chiral symmetry $\hat{\sigma}_z \hat{H}(k) \hat{\sigma}_z = -\hat{H}(k)$, $\hat{H}(k)$ has a quantised Zak phase~\cite{atala2013},
\be 
\nu = i \int_{-\pi}^{\pi} dk \langle u(k) \mid \partial_k \mid u(k) \rangle,
\ee
where $\mid u(k) \rangle$ is the Bloch function of the lower band.  When $J_2 > J_1$ the system is in the topological phase with $\nu = \pi$, hosting zero energy edge states. When $J_1 > J_2$ the system is in the trivial phase, $\nu = 0$. In finite lattices the Zak phase can be recast into the discretised form~\cite{book}
\be 
\nu = \mathrm{Im} \ln \left( \mathrm{Tr} \left[\prod_{n=1}^L \hat{P}(k_n)\right]\right), \label{eq:discrete_zak}
\ee
where $k_n = 2 \pi n / L$ is the discretised momentum space.

For simplicity, we will assume all sites have the same environmental coupling,
\be 
\hat{\Gamma} = \varepsilon \sum_{n=1}^L \left( \hat{a}_n^{\dagger} \hat{a}_n + \hat{b}_n^{\dagger} \hat{b}_n \right),
\ee
describing e.g. an array of waveguides with radiative losses occurring transverse to the lattice axis. 

\begin{figure}

\includegraphics[width=\columnwidth]{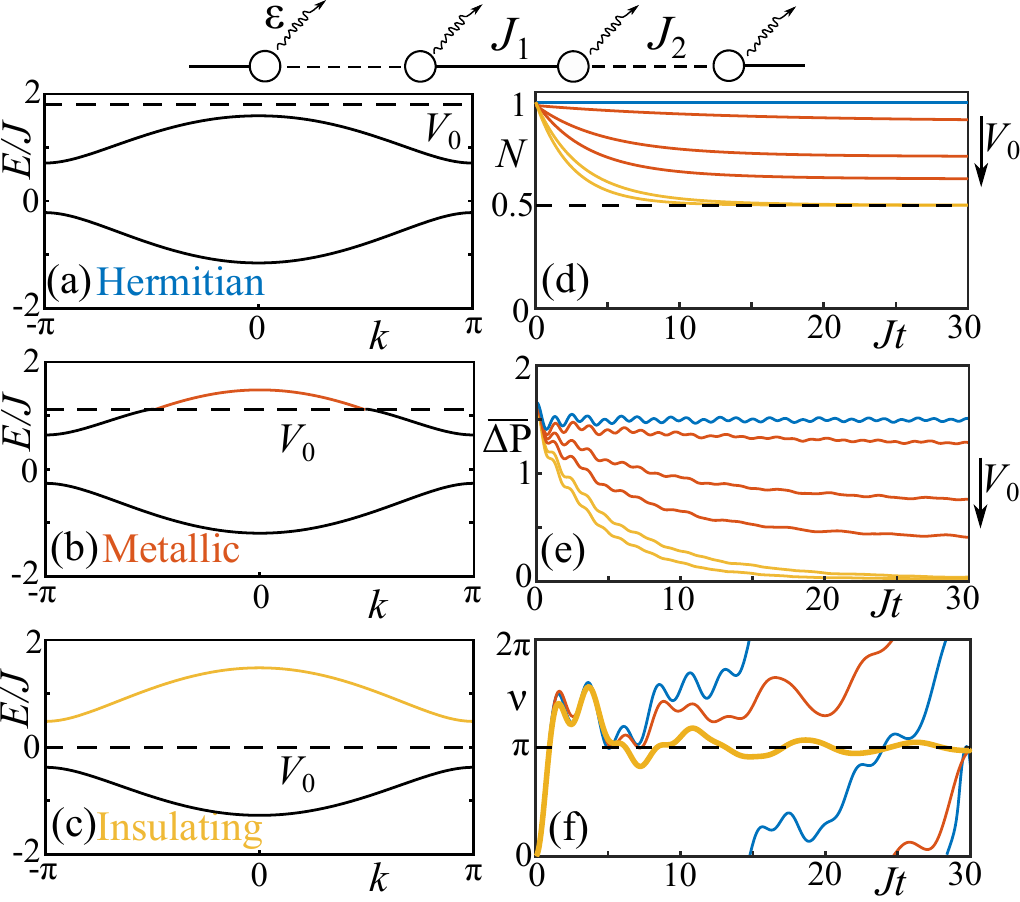}

\caption{Leaky Su-Schrieffer-Heeger lattice. (a,b,c) Bloch wave spectra for environmental potentials $V_0$ (dashed line) in three regimes: ``Hermitian'', ''metallic'', and ''insulating''. Coloured lines indicate leaky modes above $V_0$ with nonzero loss. (d) Time evolution of the total wavepacket norm $N$, which is conserved in the Hermitian limit (blue line), decays to a $V_0$-dependent limiting value in the metallic regime (brown curves), and to $50\%$ in the insulating regime (organge curves). (e) Mean error in the band projection operator estimated from the wavepacket projector approaches zero in the insulating regime. (f) Zak phase measured using the extracted band projection operator converges to a quantised limiting value in the insulating regime, even in the presence of coupling disorder. We use parameters $J_1 = J/2, J_2 = J, \varepsilon = 0.2 J$.}

\label{fig:regimes}

\end{figure}

The spectrum of the quasi-normal mode eigenvalue problem Eq.~\eqref{eq:leakyH} depends on the depth of the external potential $V_0$, as shown in Fig.~\ref{fig:regimes}(a-c). When $V_0$ exceeds the largest eigenvalue of $\hat{H}$ ($V_0 > J_1 + J_2$) all modes of the array are bound and have real energies, such that the system is effectively Hermitian. As $V_0$ is decreased, Bloch modes start to become leaky, radiating their energy into the environment and acquiring finite lifetimes. When $|J_1 -J_2| < V_0 < J_1 + J_2$ the transition between leaky and bound modes lies within the upper Bloch band of the array, and thus the number of bound modes is sensitive to the precise value of $V_0$, analogous to a metallic electronic phase. On the other hand, when $V_0$ lies in the bulk band gap, $|V_0| < |J_1 - J_2|$, the number of bound modes is independent of $V_0$, analogous to an electronic insulating phase. 

To show these different regimes exhibit different dynamical properties we simulate the time evolution of an initial state at $t=0$ via projection onto the lattice's quasi-normal modes. For simplicity, we consider a single site excitation in the bulk of a large $(L=64)$ SSH lattice, such that edge effects can be neglected. Due to the chiral symmetry of the SSH model, this single site input excites both bands uniformly, which conveniently enables measurement of the band projector $\hat{P}(k)$ without requiring averaging $\hat{F}$ over $a$ and $b$ sublattice single site excitations.

Fig.~\ref{fig:regimes}(d) shows the time evolution of the wavepacket norm for various $V_0$,
\be 
N(t) = \sum_{n=1}^L (|\psi_{a,n}(t)|^2 + |\psi_{b,n}(t)|^2),
\ee
where the initial state is normalised to $N(0)=1$. In the Hermitian regime the norm remains conserved, while in the metallic regime $N$ decays to a limiting value sensitive to $V_0$. In the insulating regime $N(t) \rightarrow 1/2$; the upper band radiates all of its energy, leaving only the modes of the lower band.

Next, Fig.~\ref{fig:regimes}(e) shows that in the insulating regime the field projection operator $\hat{F}(k,t)$ can be used to measure the band projection operator $\hat{P}(k)$. We quantify the accuracy of this measurement using the momentum-averaged error,
\be 
\overline{\Delta P}(t) = \frac{1}{L} \sum_{n=1}^L \left[ \sum_{i,j=a,b} |F_{ij}(k_n,t) - P_{ij}(k_n) |^2   \right]^{1/2},
\ee
and observe that the error becomes small once the upper band has depopulated. We note however that in our finite lattice $\overline{\Delta P}(t)$ does not converge exactly to zero, because after a sufficiently long time the wavepacket reflects off the array edges, such that the momentum space form of the projector $\hat{P}(k)$ (which assumes periodic boundary conditions) is no longer valid. We stress that in finite systems the real space form of $\hat{P}$ remains well-defined, and one can still measure quantised topological invariants~\cite{bianco2011,ringel2011,review}.

Finally, we calculate in Fig.~\ref{fig:regimes}(f) the Zak phase $\nu$ using the field projection operator, i.e. Eqs.~\eqref{eq:projector_measurement} and~\eqref{eq:discrete_zak}, including weak disorder in the coupling coefficients $\delta J \in [-W/2,W/2]$ with $W = 0.1J$. The Zak phase measured via $\hat{F}(k)$ remains robust in the insulating regime, converging to its quantised value of 0 or $\pi$ depending on whether the system is trivial or nontrivial; as the field spreads through the lattice it effectively averages over different local values of the disorder. As long as the disorder is weak enough (i.e. sufficiently long Anderson localisation length) to achieve this averaging, the result will be a good approximation to $\nu$. For stronger disorders one can still accurately measure $\nu$ by averaging over multiple input positions~\cite{bianco2011}. In contrast, previous approaches to measure the Zak phase based on the mean wavepacket displacement~\cite{rudner,PT_lattice,quantum_SSH,stjean2020} no longer work in disordered lattices, because the displacement is sensitive to the local disorder potential near the input lattice site.

For this demonstration we used a large lattice to illustrate the convergence of our method to the exact quantised Zak phase. While this method can still be applied to small arrays, there are two additional sources of error which can reduce the accuracy of Eq.~\eqref{eq:projector_measurement}: (1) Overlap of the initial excitation with edge states, if they exist, and (2) reflections off the edge, which can redistribute energy between different $k$, spoiling the completeness relation. For the parameters used in Fig.~\ref{fig:regimes} we have observed convergence to the correct Zak phase for modest system sizes of $L>6$ unit cells.

\section{Chern number in the Haldane model}
\label{sec:haldane}

\begin{figure}

\includegraphics[width=\columnwidth]{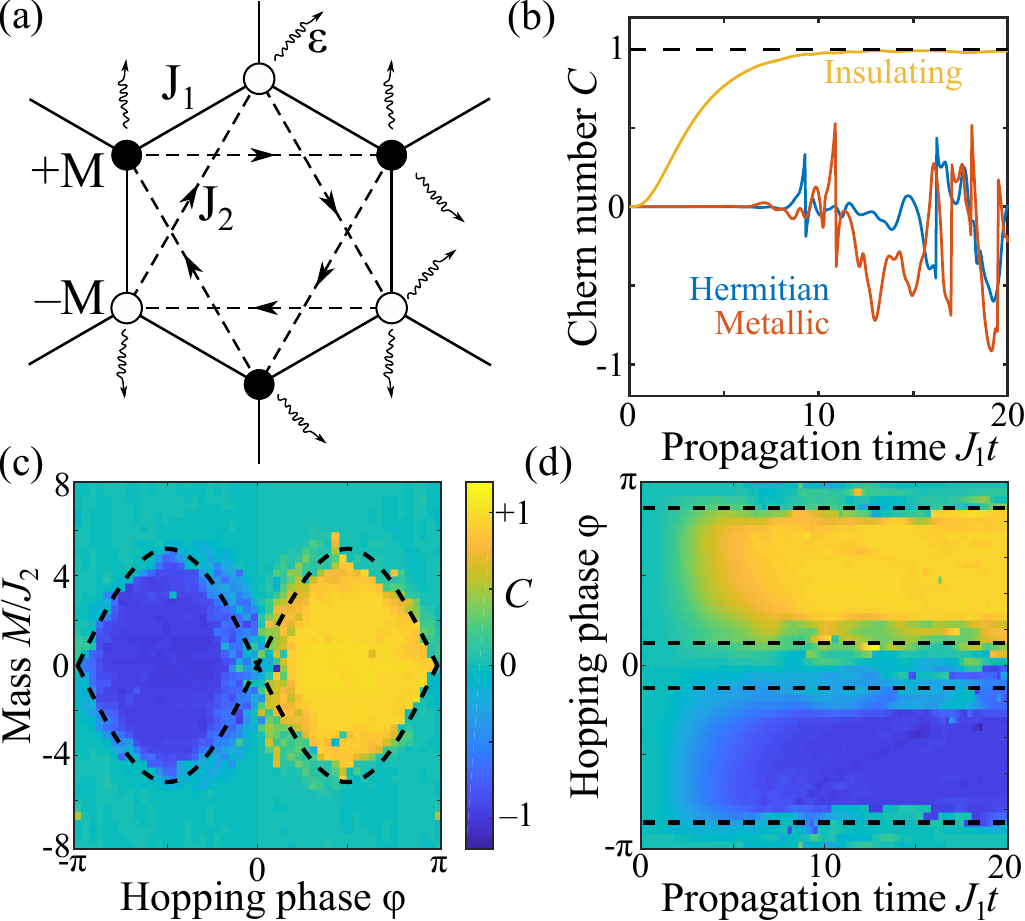}

\caption{Measurement of Chern numbers using the leaky Haldane model. (a) Schematic of the Haldane model. (b) Dynamics of the extracted Chern number in the three different regimes (Hermitian $V_0 = 4J_1$, metallic $V_0 = 2J_2$, insulating $V_0 = 0$) for $M=0$, $\varphi = \pi/2$. (c) Phase diagram extracted from the ``insulating'' lattice. (d) Dynamics of the extracted Chern number in lattices with disorder $\in [-W/2,W/2]$ in the on-site potentials, $W=M=2J_2$. Dashed lines in (c,d) indicate the exact phase boundaries of the Hermitian tight binding Hamiltonian.}

\label{fig:haldane}

\end{figure}

The Su-Schrieffer-Heeger model is a convenient testbed for exploring the measurement of bulk topological invariants, but is somewhat pathological because the value of the bulk topological invariant depends on the arbitrary choice of unit cell boundary. In this Section we show that leaky lattices can also be used to measure two-dimensional topological invariants such as the Chern number, which remains quantised even in the absence of sublattice symmetries~\cite{review}. The Haldane model is schematically illustrated in Fig.~\ref{fig:haldane}(a) and described by the tight binding Hamiltonian~\cite{haldane}
\begin{align}
\hat{H} = M \sum_{n} (& \hat{a}^{\dagger}_{n} \hat{a}_{n} -\hat{b}^{\dagger}_{n} \hat{b}_{n}) + J_1 \sum_{<n,m>} (\hat{a}_{n}^{\dagger} \hat{b}_m + \hat{b}_n^{\dagger} \hat{a}_m) \nonumber \\ &+ J_2 \sum_{\ll n,m\gg} ( \hat{a}_n^{\dagger} \hat{a}_m e^{i \varphi_{nm} } + \hat{b}_n^{\dagger} \hat{b}_m e^{-i \varphi_{nm}} ), \label{eq:haldane}
\end{align}
where $M$ is a detuning between the $a$ and $b$ sublattice depths, $J_1, J_2$ are nearest and next-nearest neighbour hopping strengths respectively, and flux sign $\varphi_{jk} = \pm \varphi$ alternates between adjacent next-nearest neighbours. Fourier transforming Eq.~\eqref{eq:haldane} yields the Bloch Hamiltonian,
\begin{align}
\hat{H}(\bs{k}) = &2 J_2 \cos \varphi \sum_i \cos (\bs{k}\cdot \bs{a}_i) \hat{\sigma}_0 + J_1 \sum_i \left[ \cos (\bs{k}\cdot \bs{\delta}_i) \hat{\sigma}_x \right. \nonumber \\ & \; \left. + \sin(\bs{k}\cdot \bs{\delta}_i) \hat{\sigma}_y \right] + [M + 2 J_1 \sin \varphi \sum_i \sin (\bs{k}\cdot \bs{a}_j) ] \hat{\sigma}_z,
\end{align}
where $\bs{\delta}_{1,2,3}$ and $\bs{a}_{1,2,3}$ are displacements between neighbouring lattice sites and unit cells respectively, and $\hat{\sigma}_n$ are Pauli matrices. The Chern number can be expressed in terms of the band projection operator $\hat{P}(\bs{k})$ as~\cite{review}
\be 
C = \frac{1}{2\pi i} \int_{BZ} \mathrm{Tr} \left[ \hat{P}(\bs{k}) [ \partial_{k_x} \hat{P}(\bs{k}), \partial_{k_y} \hat{P}(\bs{k}) ] \right] d\bs{k}. \label{eq:chern_number}
\ee
This formula for the Chern number involves the commutator of $\bs{k}$-space derivatives of the projection operators, which may be difficult to measure in practice. Luckily, Eq.~\eqref{eq:chern_number} can be efficiently discretised by replacing the $\bs{k}$ space derivatives with products over plaquettes~\cite{discretised}. In the case of the two band Haldane model, Eq.~\eqref{eq:chern_number} is discretised to
\be 
C = \frac{1}{2\pi i} \sum_{n,m} \ln \left(\mathrm{Tr} [\hat{P}_{n,m} \hat{P}_{n+1,m}\hat{P}_{n+1,m+1} \hat{P}_{n,m+1}] \right), \label{eq:chern_number_discrete}
\ee
where $\hat{P}_{n,m} = \hat{P}(\bs{k}_{n,m})$ is the band projection operator measured on a discrete grid ($n,m$) in $\bs{k}$ space. Even relatively coarse grids (lattice width $\sim 10$ unit cells) are usually sufficient to accurately measure $C$~\cite{discretised}.

For our numerical demonstration, we again assume that radiative losses occur transverse to the lattice, such that environmental coupling is uniform for all sites,
\be 
\hat{\Gamma} = \varepsilon \sum_n ( \hat{a}^{\dagger}_{n} \hat{a}_{n} + \hat{b}^{\dagger}_{n} \hat{b}_{n}).
\ee
We note that environmental coupling localised to the edges will not qualitatively affect our results; for small lattices the loss rate of the leaky bulk modes will be reduced, but without significantly affecting their cutoff energies. 

We take a finite lattice of width 14 unit cells along each principal axis, with $J_1 = 1$ and $\varepsilon = 0.2$. Single site excitations of the $a$ and $b$ sublattices at the center of the lattice are evolved up to a maximum time $t=20$ by projecting onto the quasinormal modes. We then obtain $\hat{P}$ using Eq.~\eqref{eq:projector_measurement} and the Fourier transforms of the final field profiles from each excitation. Finally, we compute the Chern number using Eq.~\eqref{eq:chern_number_discrete}. Fig.~\ref{fig:haldane}(b) shows the dynamics of the extracted Chern number $C$ in the three different regimes (Hermitian, metallic, and insulating). $C$ only converges to the correct quantised value in the insulating regime; otherwise the measured field profiles do not yield the band projection operator, due to persistent interference between the two bands.

In Fig.~\ref{fig:haldane}(c) we show the full Haldane model phase diagram extracted from $\hat{F}(\bs{k},t)$ in the insulating regime. Deep in the gapped phases the correct Chern numbers $C = 0, \pm 1$ are reproduced, while errors appear close to the phase boundaries. The two main sources of error are: (1) When the gap is small, modes near the upper band edge decay slowly and may have some residual population at the measurment time, spoiling the measurement of $\hat{P}$. (2) If the Berry curvature is strongly localised within the Brillouin zone, it may not be faithfully captured by the discretised Fourier space grid. For our choice of parameters, the latter dominates; thus the error is reduced by increasing the lattice size. Fig.~\ref{fig:haldane}(d) demonstrates that the extracted Chern number is robust against moderate on-site disorder described by $\hat{V} = \sum_n (V_{a,n} \hat{a}^{\dagger}_n \hat{a}_n + V_{b,n} \hat{b}^{\dagger}_n \hat{b}_n)$, with $V_{a,n},V_{b,n}$ uniformly distributed in $[-W/2,W/2]$. 

\section{Implementation using slab waveguides}
\label{sec:slab}

In this Section we will demonstrate that the measurement of bulk topological invariants using leaky lattices is not limited to tight binding models or site-independent environmental coupling. We consider the propagation of transverse electric (TE) polarised optical beams $\psi(x,z)$ in a one-dimensional slab waveguide array, governed by the Helmholtz equation
\be 
\partial_z^2 \psi(x,z) = -\left[\partial_x^2 + k_0^2 n^2(x) \right] \psi(x,z), \label{eq:helmholtz}
\ee 
where the refractive index profile $n(x)$ plays the role of the potential. 

\begin{figure*}

\includegraphics[width=2\columnwidth]{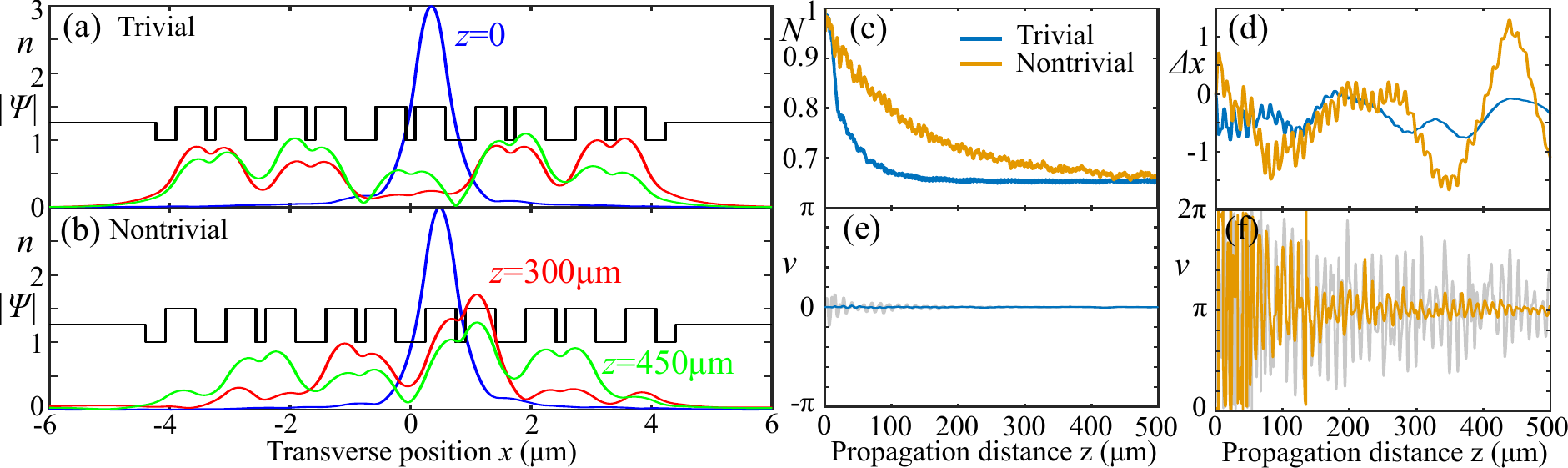}
    \caption{Array of 10 slab waveguides implementing the Su-Schrieffer-Heeger model. (a,b) Refractive index profile (black) and electric field profile $|\psi|$ after various propagation distances (coloured curves) in trivial (a, $\delta = -0.17\mu$m) and nontrivial (b, $\delta = 0.17\mu$m) lattices. (c-f) Dynamics of various observables: (c) Total norm of the electric field $N$. (d) Mean wavepacket displacement $\overline{\Delta x}$, in units of the lattice period. (e,f) Zak phase obtained from the field projection operator. Grey curves show the result of a single waveguide excitation, while coloured curves use the field projection operator obtained by averaging over initial excitations of the 5th and 6th waveguides.}
    \label{fig:SSH_slab_dynamics}
\end{figure*}

For concreteness we consider parameters similar to those used in Ref.~\cite{silicon}:  wavelength $\lambda = 1.55\mu$m, waveguides of width $w = 0.5\mu$m with core refractive index $n_{\mathrm{co}} = 1.5$, separated by a cladding with mean width $d = 0.33\mu$m and index $n_{\mathrm{cl}} = 1$. We stagger the cladding widths by $\delta = \pm 0.17\mu$m to create the $L =10$-waveguide trivial and nontrivial Su-Schrieffer-Heeger lattices shown in Fig.~\ref{fig:SSH_slab_dynamics}(a,b), and set the environmental index $n_0 = 1.26$ to lie in its bulk band gap. 

We simulate full-wave propagation dynamics in 2D by using a finite-element-method solver in COMSOL Multiphysics. To implement radiation losses, perfectly matched layers are imposed at the transverse edges of the simulation domain.
We include coupling disorder by introducing small $z$-independent variations in the waveguide separations (up to 5\% of the minimum spacing). Figure~\ref{fig:SSH_slab_dynamics}(a,b) shows the evolution of an initially-localised beam's electric field profile, which spreads throughout the entire array and radiates energy into the environment. 

The imaginary part of the array modes' effective indices $\mathrm{Im}(n_{\mathrm{eff}}) \gtrsim 10^{-3}$ corresponds to the leaky modes decaying within a propagation distance of $z \lesssim 400\mu$m, resulting in convergence of the total power in the array to a limiting value in Fig.~\ref{fig:SSH_slab_dynamics}(c). Since the environmental coupling occurs only via the ends of the array, the presence of edge states in the nontrivial phase increases the lifetime of its bulk modes, which explains the slower convergence observed. Nevertheless, the trivial and nontrivial arrays both converge to the same final power. Note that due to the strong overlap between the different waveguide modes, the power does not converge to the ideal value of 50\% observed in the tight binding model simulations of Sec.~\ref{sec:SSH}.

Before turning to the calculation of the Zak phase using the field projection operator, we first discuss the behaviour of the mean wavepacket displacement $\overline{\Delta x}(t)$,
\be 
\overline{\Delta x}(t) = x(t) - x(0),
\ee 
where
\be 
\quad x(t) = \frac{1}{2N(t)}\sum_{j=1}^L  j |\psi_j(z)|^2,
\ee 
is the beam centre of mass (normalised to the array period), and $\psi_j$ is the field amplitude in the $j$th waveguide. $\overline{\Delta x}(t)$ was previously used to extract the Zak phase, by using either loss localised on one of the sublattices~\cite{rudner,PT_lattice}, measuring its time average~\cite{quantum_SSH}, or performing energy-resolved measurements~\cite{stjean2020}.

Due to the coupling disorder the wavepacket displacement plotted in Fig.~\ref{fig:SSH_slab_dynamics}(d) does not oscillate about or converge to any quantised value, and thus the field projection operator is required to determine the Zak phase of the lattice. 

Fig.~\ref{fig:SSH_slab_dynamics}(e) reveals a rapid convergence to the correct Zak phase in the trivial phase, even without averaging the field projector over excitations of the two sublattices. On the other hand, the nontrivial array exhibits a slower convergence to the correct Zak phase in Fig.~\ref{fig:SSH_slab_dynamics}(f), which we attribute to its lower losses and our small system size. These numerical results establish that our theory remains valid even for very small systems, and without requiring the tight binding approximation.

\section{Discussion and Conclusion}
\label{sec:conclusion}

We have shown how non-interacting waves propagating in shallow lattices can be described by a novel class of tight binding models with energy-dependent losses. Different propagation regimes emerge depending on the energy detuning between the lattice and its environment: ``Hermitian'' (lossless), ``metallic'' (sensitive to the detuning), and ``insulating'' (insensitive to the detuning). At long propagation times the field profile in the insulating regime allows measurement of the band projection operator and quantised topological invariants. Leaky lattices are more flexible than existing approaches for measuring topological invariants using propagation dynamics, which either exploit known symmetries of the Hamiltonian~\cite{longhi2019}, or require time-consuming band tomography using a spatially-broad initial wavepacket tailored to excite a single band~\cite{wimmer2017}.

We validated our scheme using the Su-Schrieffer-Heeger and Haldane models and numerical simulations of light propagation in a slab waveguide array, demonstrating that bulk topological invariants can still be measured in small and disordered lattices. As further verification, Appendix~\ref{sec:continuum} contains a detailed discussion of quasi-normal modes of the continuum Schr\"odinger equation, and Appendix~\ref{sec:coupler} shows that the discrete tight binding models we have analyzed can be a good description of shallow (strongly coupled) waveguides described by the continuum Schr\"odinger equation.

Our models can also be implemented using existing platforms for topological lattices, such as deep waveguide arrays or photonic crystals. In the former, one can introduce an environment using auxiliary waveguide arrays weakly coupled to the sites of the lattice of interest, similar to the approach experimentally demonstrated in Ref.~\cite{mukherjee}. For the latter, one could embed the topological photonic crystal within an appropriately-designed photonic crystal environment~\cite{cerjan}. We note that while our approach requires knowledge of both the intensity and phase of the field, one can determine the phase by measuring the intensity in real space and Fourier space~\cite{GS}.

To simplify our presentation we focused on the measurement of topological invariants in Fourier space, which assumes translation invariance. However, according to Refs.~\cite{bianco2011,ringel2011,review}, one can equivalently use real space topological markers to obtain bulk topological invariants using correlation functions of the field. An interesting future direction is to use leaky photonic lattices to study the real space dynamics of correlation functions such as the Chern marker~\cite{caio2019}. 

In the insulating regime, when the detuning of the leaky bands from the cutoff $V_0$ is much larger than the environmental coupling strength, the modal losses $\gamma_n$ become approximately proportional to their energy detuning, $\gamma_n =  (V_0 - E_n)\alpha$, where $\alpha$ is some constant factor. Consequently, at finite times (before the leaky bands have completely depopulated), the magnitude of the band weights in Eq.~\eqref{eq:f1} will resemble Boltzman factors with effective temperature $T = 1/(\alpha t)$, $|c_n| \propto e^{-(V_0 - E_n)/T}$. Thus, leaky lattices may also provide a platform to emulate topological systems at finite temperature~\cite{viyuela2014}.

Finally, leaky lattices are also a highly promising platform for studying related topics such as non-Hermitian systems and non-Hermitian topological phases. Our models based on the non-interacting Schr\"odinger equation are applicable to various bosonic wave systems including light propagation in waveguide arrays and photonic crystals, as well as acoustics and Bose-Einstein condensates, where the phenomena we have discussed may be further enriched by effects such as spin-orbit coupling and inter-particle interactions.

\section*{Acknowledgements}

We thank Alexander Cerjan, Zhigang Chen, Yidong Chong, and Mikael Rechstman for illuminating discussions. D.~L. is supported by the Institute for Basic Science in Korea (IBS-R024-Y1). D.~A.~S. acknowledges funding from the Australian Research Council Early Career Researcher Award (DE190100430) and the Russian Foundation for Basic Research (Grant No. 18-02-00381). 

\appendix

\section{Variational formulation of leaky mode eigenvalue problems}
\label{sec:continuum} 

In this Appendix we provide further details of the derivation of Eq.~\eqref{eq:leakyH} in the main text. Our starting point is the time-independent Schr\"odinger euqation for propagation-invariant eigenmode profiles $\psi(\bs{r},t) = \phi(\bs{r}) e^{-i E t}$,
\be 
E \phi(\bs{r}) = [-\frac{1}{2m}\nabla^2 + V(\bs{r})] \phi(\bs{r}), \label{eq:TISE}
\ee
which forms a Hermitian eigenvalue problem. When $V(\bs{r})$ takes a constant value $V_0$ outside the finite lattice, modes are conventionally divided into bound states with energies $E < V_0$ and a continuum delocalised scattering states with $E > V_0$. Together these form a complete basis for describing the dynamics of the lattice and its environment. However, this basis cannot provide a concise intuitive description of the dynamics of states initially localised to $V(\bs{r})$ but with energies $E > V_0$, which must be expressed as a superposition of the continuum of scattering states. In this case, quasi-normal modes provide a very useful tool for understanding the dynamics~\cite{quasinormal_review}.

Quasi-normal modes are shape-invariant solutions of the Schr\"odinger equation under outgoing wave boundary conditions. For example, consider a one-dimensional system for which $V(x)$ vanishes outside the bounded domain $x \in (0, L)$. We seek solutions of the form $\psi(x,t) = \phi (x) e^{-i (E + i \gamma) t}$, where $E$ is the energy and $\gamma$ is the modal growth rate. Outside the bounded domain the modal profile is $\phi(x) \propto \exp(\pm \xi x)$, where
\be 
\xi = \pm \sqrt{2m(V_0 - E - i \gamma)}. ~\label{eq:xi}
\ee
Inside the domain the modal profile $\phi(x)$ satisfies
\begin{subequations}
\label{eq:robin}
\begin{align}
&(E+i\gamma) \phi + \frac{1}{2m}\partial_x^2 \phi - V(x) \phi = 0, \quad x \in (0,L) \label{eq:eigenvalue} \\
&\partial_x \phi(0) - \xi \phi(0) = 0, \label{eq:b1} \\
&\partial_x \phi(L) + \xi \phi(L) = 0. \label{eq:b2}
\end{align}
\end{subequations}
Eqs.~\eqref{eq:robin} can be solved using the variational method. Namely, expanding the modal profile $\phi(x)$ in terms of a basis set $\phi(x) = \sum_n c_n \phi_n$ yields the nonlinear eigenvalue problem
\begin{subequations}
\label{eq:nonlinear_matrix}
\begin{align}
& \mathrm{det}\left( \hat{H} + \xi \hat{\Gamma} + (\xi^2 - V_0) \hat{S} \right) = 0, \\
& H_{mn} = \int_0^L \left[ V(x) \phi_m^* \phi_n + \frac{1}{2m}(\partial_x \phi_m^*)(\partial_x \phi_n) \right] dx, \\
& \Gamma_{mn} = \phi_m^*(L) \phi_n(L) + \phi_m^*(0) \phi_n(0), \\
& S_{mn} = \int_0^L \phi_m^* \phi_n dx.
\end{align}
\end{subequations}
The first term ($\hat{H}$) forms an effective Hamiltonian, corresponding to matrix elements of the energy operator $-\frac{1}{2m}\partial_x^2 + V(x)$. The last term ($\hat{S}$) accounts for the overlap between different basis elements. The second term ($\hat{\Gamma}$) describes the coupling of the basis states to the environment (domain boundary) and makes the problem effectively non-Hermitian. 

Eqs.~\eqref{eq:nonlinear_matrix} can be discretised using the finite element method (see e.g. Ref.~\cite{leaky_variational}), which yields a sparse discrete nonlinear eigenvalue problem solvable using standard numerical methods. Higher accuracy solutions can then be obtained by applying Newton's method to the original boundary value problem~\cite{leaky_variational}.

One method of solving Eqs.~\eqref{eq:nonlinear_matrix} is linearisation, based on introducing the auxiliary field $\Phi = (\xi \phi, \phi)$~\cite{QEP}. $\Phi$, and hence $\phi$, can be obtained by solving the linear non-Hermitian generalised eigenvalue problem
\be 
\left( \begin{array}{cc} \hat{\Gamma} & \hat{H} - V_0 \hat{S} \\ -\hat{1} & 0 \end{array} \right) \Phi = -\xi \left( \begin{array}{cc} \hat{S} & 0 \\ 0 & \hat{1} \end{array} \right) \Phi. \label{eq:linearised}
\ee
Given the eigenvalues $\xi$ one can readily obtain the modal energies $E$ and growth rates $\gamma$ using Eq.~\eqref{eq:xi}. Modes with $E < V_0$ are bound states of the potential and have purely real eigenvalues, $\gamma = 0$, corresponding to purely real $\xi$. When $E > V_0$ the modes are divided into scattering states ($\gamma = 0$ and $\xi$ purely imaginary) and quasi-normal modes ($\gamma \ne 0 $ and $\xi$ complex). The latter are further divided into leaky modes that decay by emitting radiation ($\gamma < 0$) and growing modes that absorb radiation ($\gamma > 0$); this doubling of modes occurs because the dimension of Eq.~\eqref{eq:linearised} is twice that of the linear eigenvalue problem obtained in the limit $\hat{\Gamma} \rightarrow 0$ of Eq.~\eqref{eq:nonlinear_matrix}.

\section{Leaky coupled mode theories and tight binding models}
\label{sec:coupler}

To obtain analytical insight into the quasi-normal modes of leaky lattices it is useful to consider approximate solutions of Eq.~\eqref{eq:nonlinear_matrix}, assuming weak coupling between different lattice sites and to the environment. 

To obtain coupled mode theories one can use modes of the individual lattice sites as a basis for Eq.~\eqref{eq:nonlinear_matrix}. That is, $V(x) = V_{\mathrm{cl}} + \sum_n V_n (x)$ is decomposed into its individual sites, where $V_{\mathrm{cl}}$ is a uniform background potential (note this is distinct from the environmental potential $V_0$), and $V_n(x)$ vanishes outside the $n$th site. Then, solving Eq.~\eqref{eq:nonlinear_matrix} numerically with $V(x) \rightarrow V_n(x)+V_{\mathrm{cl}}$ one can obtain a basis for the coupled mode theory. Diagonal elements of $\hat{H}$ will correspond to site energies, with off-diagonal elements describing coupling between different sites. The modal overlaps described by off-diagonal elements of $\hat{S}$ may be significant, such that Eq.~\eqref{eq:linearised} forms a generalised eigenvalue problem.

If we further assume that the overlap between different basis elements is negligible, such that $\hat{S} \rightarrow \hat{1}$, we obtain non-Hermitian ``tight binding'' models of the form
\be 
\left( \begin{array}{cc} \hat{\Gamma} & \hat{H} - V_0 \hat{1}  \\ -\hat{1} & 0 \end{array} \right) \Phi = -\xi \Phi. \label{eq:linearised2}
\ee
Note that in contrast to regular non-Hermitian eigenvalue problems, the eigenvalue $\xi$ is not the mode energy; $V_0 - \xi^2/(2m)$ is. Consequently, the mode energies remain purely real until they cross $V_0$.

\begin{figure}

\includegraphics[width=\columnwidth]{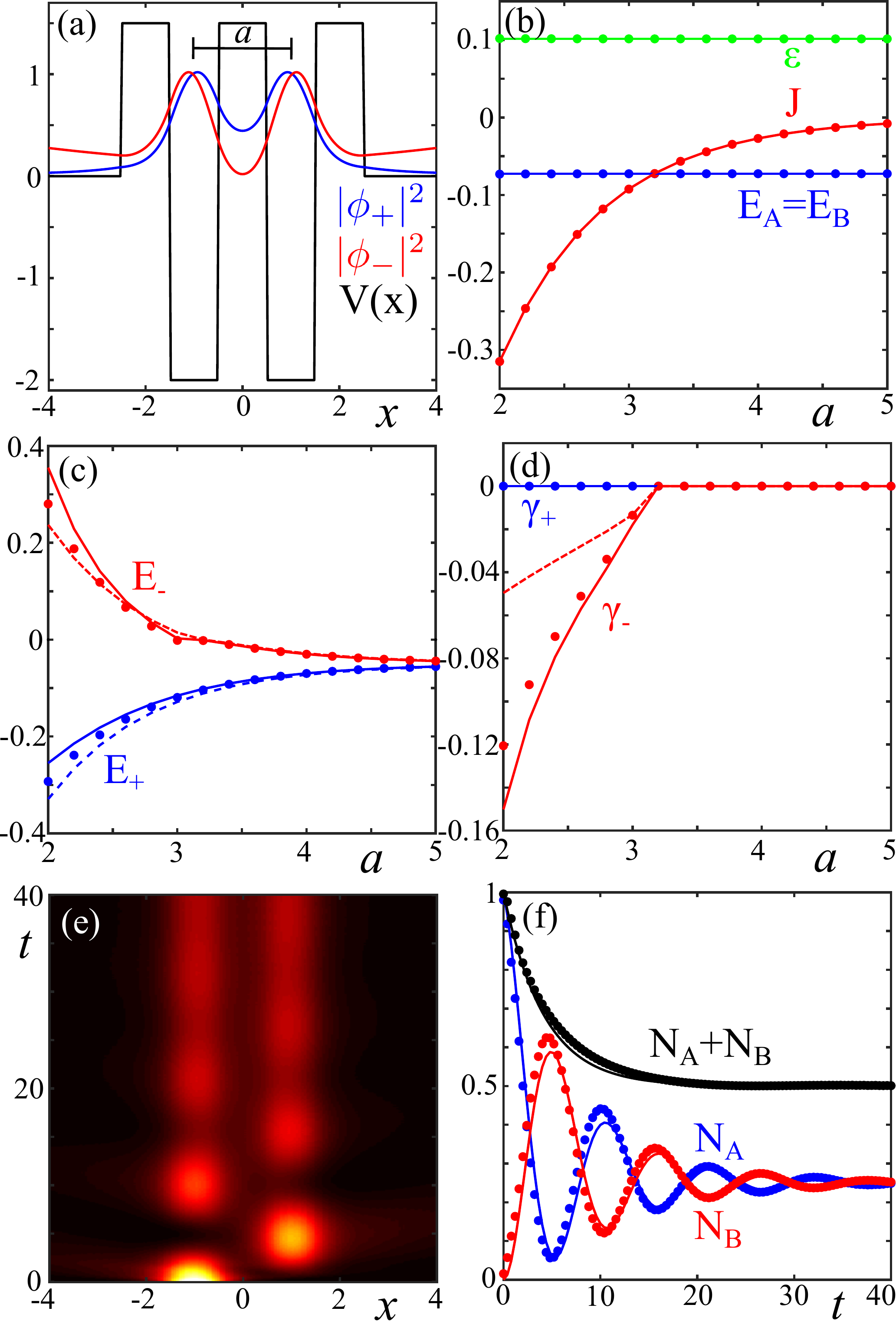}

\caption{(a) Leaky dimer potential and and intensity profiles of its fundamental symmetric $\phi_+$ and antisymmetric $\phi_-$ modes. (b) Tight binding model parameters in Eq.~\eqref{eq:dimer} as a function of the waveguide separation $a$ (solid lines) fit to numerically-obtained coupled mode theory parameters (circles). (c,d) Modal energies (c) and decay rates (d) as a function of the well separation. Solid lines are obtained from the numerical coupled mode theory (see text for details), while dashed lines are obtained from the tight binding model. Circles indicate the direct numerical solution of Eqs.~\eqref{eq:nonlinear_matrix} (e) Field intensity evolution calculated using beam propagation simulations of the Schr\"odinger equation with absorbing boundary conditions and $a=2$. (f) Evolution of the power in the individual waveguides (red,blue) and total power (black). Solid lines are obtained from the tight binding model Eq.~\eqref{eq:dimer}.}

\label{fig:dimer}

\end{figure}

To establish that these simplified coupled mode theories and tight binding models can adequately describe real systems, we consider the simplest nontrivial example of a dimer formed by two identical square wells separated by a variable distance $a$, shown in Fig.~\ref{fig:dimer}(a). Here, lengths are normalized by the wavenumber such that one can fix $m=1$. The well depths are chosen so that each hosts a single bound mode with $E < V_0 = 0$ when they are well-separated. As the well separation is reduced the single well modes hybridize into symmetric and anti-symmetric supermodes; the latter's energy crosses $V_0$ and becomes leaky. 

We compare three methods to solve Eqs.~\eqref{eq:nonlinear_matrix}: the numerically-exact finite element method, the coupled mode theory obtained by decomposing $V(x)$ into its constituent wells and using their individual modes as a basis, and the analytically-solvable tight binding model
\be 
\hat{H} = \left(\begin{array}{cc} E_0 & J \\ J & E_0 \end{array} \right), \quad \hat{\Gamma} = \left(\begin{array}{cc} \varepsilon & 0 \\ 0 & \varepsilon \end{array}\right), \quad  \hat{S} = \left( \begin{array}{cc} 1 & 0 \\ 0 & 1 \end{array} \right), \label{eq:dimer}
\ee 
where the coupling $J \approx 3.66 \exp(-1.23 a)$, and parameters $E_0, \varepsilon$ are independent of the well separation, to a good approximation. Fig.~\ref{fig:dimer}(b) compares the tight binding model parameters against the exact matrix elements of the coupled mode theory; the agreement is very good. We note, however, that the tight binding model neglects off-diagonal elements of $\hat{S}$, which approach 20\% for small separations. 

The solution of the eigenvalue problem Eq.~\eqref{eq:linearised2} is
\be 
\xi =  -\frac{\varepsilon}{2} \pm \sqrt{ \pm J -  E_0 + \varepsilon^2/4}, \label{fig:dimer_eigenvalues}
\ee
where all four combinations of $\pm$ should be considered. Due to the symmetry of the dimer, the eigenstates must be either symmetric $\mid \phi_+ \rangle = (1,1)/\sqrt{2}$ or antisymmetric, $\mid \phi_- \rangle = (1,-1)/\sqrt{2}$. The two eigenvalues of interest (corresponding to bound or leaky modes) are
\begin{align}
E_+ &= E_0 + J - \frac{\varepsilon^2}{2} + \varepsilon \sqrt{ \varepsilon^2/4 - E_0 - J}, \\
E_- + i \gamma_- &= E_0 - J - \frac{\varepsilon^2}{2} - \varepsilon \sqrt{ \varepsilon^2/4 - E_0 + J}.
\end{align}
Fig.~\ref{fig:dimer}(c,d) plots the modal energies and decay rates, restricting to the physical solutions that are either bound or radiate energy. We observe that the tight binding model accurately reproduces the energies, but underestimates the decay rate of the antisymmetric mode. Including the non-zero off-diagonal elements of $\hat{S}$ does not fix this discrepancy; more accurate results from the coupled mode theory require the inclusion of the next lowest energy (but unphysical) modes of the individual wells that diverge as $|x| \rightarrow \infty$. The solid lines in Figs.~\ref{fig:dimer}(c,d) are obtained from a 4 mode coupled mode theory including these anomalous modes. The agreement is much better.

Given the quasi-normal modes, one can obtain an effective Hamiltonian describing the evolution of an arbitrary initial state by projecting onto the quasi-normal modes $\mid \phi_{\pm} \rangle$, $\hat{H}_{\mathrm{eff}} = (E_+ + i \gamma_+) \mid \phi_+ \rangle \langle \phi_- \mid + (E_- + i \gamma_-) \mid \phi_- \rangle \langle \phi_- \mid$. Since $\mid \phi_{\pm} \rangle = (1,\pm 1)/\sqrt{2}$, this yields
\be 
\hat{H}_{\mathrm{eff}} = \frac{1}{2} \left(\begin{array}{cc} E_+ + E_- + i \gamma_-  & E_+ - E_- - i \gamma_- \\ E_+ - E_- - i \gamma_- & E_+ + E_- + i \gamma_- \end{array} \right).
\ee
When both modes are bound ($\gamma_{\pm} = 0$) this is just the Hamiltonian for a regular Hermitian dimer with effective coupling strength $(E_+ - E_-) / 2$. On the other hand, when one of the modes is leaky the effective Hamiltonian contains non-Hermitian diagonal and off-diagonal elements of equal magnitude $\gamma_-$, the latter providing a simple realization of non-Hermitian coupling. Previous proposals for implementing non-Hermitian coupling require either complicated arrangements of coupling via auxiliary waveguides~\cite{longhi1,longhi2,leykam2017,mukherjee}, or have the non-Hermitian coupling limited to very weak values compared to the diagonal non-Hermitian terms~\cite{lattice_loss}; here the two are equal and emerge simply by creating lattices out of single site modes close to their cutoff energy.

When both modes are bound, excitation of a single well yields persistent oscillations between the two sites. Fig.~\ref{fig:dimer}(e) shows the different behaviour occurring when one of the coupler modes is leaky. In this case, there is an initial transient oscillation between the two sites. Eventually the antisymmetric mode radiates all of its energy, leaving just an excitation of the symmetric mode and a $t$-independent intensity profile. The power within the sites during this damped oscillation, plotted in Fig.~\ref{fig:dimer}(f), is well-approximated by our simple dimer model Eq.~\eqref{eq:dimer}. Thus, arrays of shallow coupled wells provide a simple platform for realising effective tight binding models with energy-dependent losses. By making a lattice formed by these dimers one can readily obtain a continuum model of the SSH lattice of Sec.~\ref{sec:SSH}~\cite{malkova2009}.

\section{Decay of off-diagonal elements of the field projector}
\label{sec:phase}
To justify smallness of the interband terms at large times, we for definiteness consider two bands of interest in a 1D lattice with $k \in [-\pi,\pi]$. The time-dependent Fourier transform of the field is given by
\begin{multline}
\psi(k,t) = c_{+}(k) e^{-i[E_{+}(k) + i\gamma_{+} (k)]t} \ket{u_{+}(k)} \\ + c_{-}(k) e^{-iE_{-}t} \ket{u_{-}(k)}\:, 
\end{multline}
where subscripts $\pm$ refer to the upper and lower bands, respectively. Depending on the environmental potential $V_0$ the upper band can incorporate radiative losses $\gamma_{+}(k)$. 

We consider the field projection operator in real space,
\begin{equation} \label{eq:Pxt}
\hat{F}(x,t) = \int_{-\pi}^{\pi} \ket{\psi(k,t)}\bra{\psi(k,t)} e^{ikx} dk\:.
\end{equation}
If $\gamma_{+}(k)$ is nonzero throughout the upper band the off-diagonal terms of the projector decay exponentially, leaving $\hat{F}(x,t) \rightarrow \hat{P}(x)$. In the following calculations we want to show that $\hat{F}(x,t)$ still converges to the projector when there are multiple bound bands. We therefore set $\gamma_{+} (k) = 0$, e.g. assume $V_0>\text{max} E_{+} (k)$. For simplicity we assume a symmetric spectrum, $E_{+}(k) = - E_{-}(k)$. Then Eq.~\eqref{eq:Pxt} returns
\begin{multline}
\hat{F}(x,t) =  \int_{-\pi}^{\pi} |c_{+}(k)|^2 \ket{u_{+}(k)} \bra{u_{+}(k)} e^{ikx} dk \\  +\int_{-\pi}^{\pi} |c_{-}(k)|^2 \ket{u_{-}(k)} \bra{u_{-}(k)} e^{ikx} dk  \\  +\int_{-\pi}^{\pi} \left( c_{+}(k) c^{*}_{-}(k)  e^{2iE_{-}(k)t + i k x}  \ket{u_{+}(k)} \bra{u_{-}(k)} \right.\\ + \left. c^{*}_{+}(k) c_{-}(k)  e^{-2iE_{-}(k)t + i k x}  \ket{u_{-}(k)} \bra{u_{+}(k)} \right) dk
\end{multline}
Here, the first two terms reproduce the band projection operator (in real space), while the third integral $I_3$ is the error term. At large times, it can be calculated by the stationary phase method. The two stationary phase points $\pm k_{\text{st}}$ are found from the condition 
$-\beta t \pm  x = 0 $, where $\beta = -2\dfrac{d E_{-} (k_{\text{st}})} {dk }$ is the difference in the group velocities of the two bands. We expand the energy $E_{-}(k_{\text{st}} + \tilde{k}) = E_{-} (k_{\text{st}}) - \beta \tilde{k} /2 + m_{\mathrm{eff}}^{-1}  \tilde{k}^2$, denoting the second derivative $m_{\mathrm{eff}}^{-1} =  \dfrac{1}{2} \dfrac{d^2 E_{-} (k_{\text{st}})} {d k^2}$. 
In the moving coordinate frame $x = \beta t$, we obtain
\begin{multline}
I_3 = \left(  c_{+} (k_{\text{st}}) c^{*}_{-} (k_{\text{st}}) \ket{u_{+}(k_{\text{st}})} \bra{u_{-} (k_{\text{st}}) } e^{2iE_{-} (k_{\text{st}}) t}  \right. \\ + \left. c^{*}_{+}(-k_{\text{st}}) c_{-}(-k_{\text{st}})  e^{-2iE_{-} (k_{\text{st}}) t}  \ket{u_{-}((-k_{\text{st}})}\bra{u_{+} (-k_{\text{st}})}\right)\\ \times \dfrac{\sqrt{\pi m_{\mathrm{eff}}}}{\sqrt{i 2 t}}\:,
\end{multline}
Contribution from the end points $\pm \pi$ of the integration interval is variable $\propto ({i\beta t})^{-1}$.

Thus, the error term vanishes $\propto t^{-1/2}$, enabling measurement of the band projection operator in real space.

\end{document}